\documentstyle{article}
\begin{document}
\def\1{{\rm 1 \kern -.10cm I \kern .14cm}}
\def\R{{\rm R \kern -.28cm I \kern .19cm}}

\begin{titlepage} 
\begin{flushright}
  LPTHE-ORSAY 96/79 \\hep-ph/9610257
\end{flushright} 
\vskip .8cm 
\centerline{\LARGE{\bf {Neutrino mass models}}}
\vskip .5cm
\centerline{\LARGE{\bf {with an abelian family symmetry}}\footnote{To
  appear in the proceedings of the NATO ASI on Masses of Fundamental
  Particles, Carg\`ese, France, August 5-17, 1996.}}  
\vskip 1.5cm
\centerline{\bf {St\'ephane Lavignac\footnote{Supported in part by the
  Human Capital and Mobility Programme, contract CHRX-CT93-0132.}}}
\vskip .5cm
\centerline{\em Laboratoire de Physique Th\'eorique et Hautes
  Energies\footnote{Laboratoire associ\'e au CNRS-URA-D0063.}}
\centerline{\em Universit\'e Paris-Sud, B\^at. 210,}
\centerline{\em F-91405 Orsay C\'edex, France }

\vskip 2cm
\centerline{\bf {Abstract}}

\indent
Abelian family symmetries provide a predictive framework for neutrino
mass models. In seesaw models based on an abelian family symmetry, the
structures of the Dirac and the Majorana matrices are derived from the
symmetry, and the neutrino masses and mixing angles are determined by
the lepton charges under the family symmetry. Such models can lead to
mass degeneracies and large mixing angles as well as mass hierarchies,
the squared mass difference between quasi-degenerate neutrinos being
determined by the symmetry. We present two models illustrating this
approach.

\end{titlepage} 


\section{Introduction}
\indent

Fermion masses are one of the most fundamental problems of particle
physics. While the origin of the observed hierarchy between quark and
charged lepton masses remains unexplained in the Standard Model and most
of its extensions, the question of whether the neutrinos are massive or
not is still open. On the theoretical side, since neutrino masses are
not protected by any fundamental symmetry\footnote{Neutrino masses are
only protected by lepton number symmetry, which is an accidental global
symmetry of the Standard Model, and turns out to be violated in most of
its extensions.}, there is no reason to expect them to be zero. Now, if
the neutrinos are massive, the rather unnatural suppression of their
masses relative to the quarks and charged leptons of the same family has
to be explained. On the phenomenological side, massive neutrinos could
solve in a natural way several astrophysical and cosmological problems.

Family symmetries, which have been first introduced in order to explain
the quark mass and mixing hierarchies, provide a predictive framework
for neutrino mass models. We present two seesaw models based on an
abelian family symmetry, the one leading to a hierarchical mass
spectrum, the other yielding two quasi-degenerate neutrinos with a large
mixing angle.

\section{Neutrino oscillations}
\indent

There is no direct laboratory evidence for non-zero neutrino masses (the
present upper bounds are $m_{\nu_{e}} < 5.1 \; eV$, $m_{\nu_{\mu}} < 160
\; keV$, $m_{\nu_{\tau}} < 24 \; MeV$), but some  experimental data
suggest neutrino oscillations. Before discussing them, let us briefly
review neutrino oscillations in the simple case of two flavours.
Neutrinos oscillate when the weak eigenstates $\nu_{\alpha, \beta}$ are
mixtures of the mass eigenstates $\nu_{1,2}$:
\begin{equation}
  \left( \begin{array}{c}
	  	\nu_\alpha \\
		\nu_\beta
	 \end{array}  \right)  =
  \left( \begin{array}{rr}
	  	  \cos \theta  &  \sin \theta \\
		- \sin \theta  &  \cos \theta
	 \end{array}  \right)
  \left( \begin{array}{cc}
	  	\nu_1 \\
		\nu_2
	 \end{array}  \right)
\end{equation}
Suppose a weak eigenstate $\nu_\alpha$ is produced at $t=0$: $\nu(0) =
\nu_\alpha = \cos \theta \: \nu_1 + \sin \theta \: \nu_2$. At $t$ it
will have evolved into $\nu(t) = \cos \theta e^{-i E_1 t} \nu_1 + \sin
\theta e^{-i E_2 t} \nu_2$. Assuming relativistic neutrinos, the
probability of detecting the other weak eigenstate is:
\begin{equation} 
  P(\nu_\alpha \rightarrow \nu_\beta) = | < \nu_\beta | \nu(t) > |^2 =
\sin^2 2 \theta \: \sin^2 \left( \frac{\Delta m^2 L}{4 E} \right)
\end{equation}
where $E$ is the neutrino energy, $L$ the distance travelled by the
neutrino between the source and the detector, and $\Delta m^2 = m_2^2 -
m_1^2$ is the squared mass difference between the two mass eigenstates.
Note that the oscillation probability, which is a function of the ratio
$L/E$, is characterized by the parameters $\sin^2 2 \theta$ and $\Delta
m^2$.

The strongest indication in favour of neutrino oscillations comes from
the solar neutrino deficit. All solar neutrino experiments have observed
a suppression of the $\nu_e$ flux relative to the predictions of the
standard solar models. The most convincing explanation of this deficit
is provided by the Mikheyev-Smirnov-Wolfenstein (MSW) \cite{MSW}
conversion of the electron neutrino into another specie inside the sun.
There are two allowed regions in the ($\sin^2 2 \theta$, $\Delta m^2$)
plane, one with a small mixing angle ($\Delta m^2 \sim 6.10^{-6} eV^2$
and $\sin^2 2 \theta \sim 10^{-3}-10^{-2}$), the other with a large
mixing angle ($\Delta m^2 \sim 10^{-5}-10^{-4} eV^2$ and $0.2 \leq
\sin^2 2 \theta \leq 0.9$). Another hint for neutrino oscillations is
the atmospheric neutrino anomaly. Some experiments have measured a
significantly lower ratio of the $\nu_\mu$ to the $\nu_e$ flux than
predicted, which could be a signature of $\nu_\mu$ oscillations into
another flavour with a large mixing angle ($\Delta m^2 \sim 10^{-2}
eV^2$ and $\sin^2 2 \theta \geq 0.5$). However, this anomaly has not
been observed by all experiments, and some uncertainties remain.
Finally, there may be indications in favour of $\bar \nu_\mu
\rightleftharpoons \bar \nu_e$ oscillations with a small mixing angle
($\sin^2 2 \theta \sim$ a few $10^{-3}$ for $\Delta m^2$ in the few
$eV^2$ region) from the LSND experiment. This interpretation needs to be
confirmed.

Massive neutrinos are also interesting for cosmology. Structure
formation requires, in addition to cold dark matter, a small amount of
hot dark matter, which could be composed of a neutrino with mass between
$1$ and $10 \: eV$, or several degenerate neutrinos in the few $eV$
range.

\section{Models of neutrino masses}

\subsection{Generalities}
\indent

There are numerous models of neutrino mass. All of them need an
extension of the particle content of the Standard Model. A Dirac mass
term (${\cal L}_m = - \, m_D \, \bar \nu_L N_R + h.c.$) requires the
introduction of a right-handed ($RH$) neutrino $N_R$ in addition to the
standard left-handed ($LH$) neutrino $\nu_L$. Since $N_R$ is a $SU(2)_L$
singlet, such a mass term violates the weak isospin by $\Delta I_W =
1/2$, and must therefore be generated by a Yukawa coupling to a Higgs
doublet: ${\cal L}_{Yuk} = - \, h_\nu \, ( \bar \nu_L \bar e_L ) \, H_2
\, N_R + h.c. \,$, with $m_D = h_\nu <H^0_2>$. A Dirac neutrino is then
like other fermions, but its Yukawa coupling $h_\nu$ has to be
unnaturally small. A Majorana mass term (${\cal L}_m = - 1/2 \, m_M \,
\bar \nu_L \nu^c_R + h.c.$), which violates lepton number by two units
($\Delta L = 2$), involves a transition from the standard $\nu_L$ ($I_W
= + 1/2$) to its CP conjugate $\nu_R^c$ ($I_W = - 1/2$). Such a mass
term has $\Delta I_W = 1$ and must therefore originate from a Yukawa
coupling to a weak Higgs triplet (Gelmini-Roncadelli model) or from an
effective interaction.

In the presence of a $RH$ neutrino, both Dirac and Majorana mass terms
can be present, as well as a $\Delta I_W = 0$  Majorana mass term for
the $RH$ neutrino, ${\cal L}_m = - 1/2 \, M_R \, \bar N^c_L N_R + h.c.
\,$. The full mass term takes then the following form (in the absence of
a Higgs triplet):
\begin{equation}
  - {1 \over 2} \left( \bar \nu_L \bar N_L^c \right)
  \left( \begin{array}{cc}
	  	0  &  m_D \\
		m_D  &  M_R
	 \end{array}  \right)
  \left( \begin{array}{c}
		\nu_R^c  \\  N_R
	 \end{array}  \right) + h.c.
\label{eq:seesaw}
\end{equation}
The physical neutrino states, which are two Majorana neutrinos, are
obtained from the diagonalization of this 2x2 matrix. Particularly
interesting is the seesaw limit \cite{seesaw}, $m_D \ll M_R$, in which
one eigenstate has a mass far below the weak scale\footnote{The Dirac
mass $m_D$, which is protected by the electroweak symmetry, is expected
to be of the order of the breaking scale $M_{weak} = 246 \; GeV$,
whereas the Majorana mass $M_R$, being not constrained by any symmetry,
can be much larger than $M_{weak}$. Thus $m^2_D / M_R \ll M_{weak}$.}:
\begin{eqnarray}
  \left\{  \begin{array}{l}
	m_1 \simeq m^2_D / M_R  \\
	m_2 \simeq M_R
  \end{array}  \right.
  &  \mbox{and}  &  \tan \theta \simeq m_D / M_R
\end{eqnarray}
Since the mixing angle is small, the light eigenstate is mainly the
Standard Model neutrino.

\subsection{Seesaw models}
\indent

The seesaw mechanism is very popular, because it naturally generates
neutrino masses much lighter than the weak scale. Moreover, it can be
easily implemented in numerous extensions of the Standard Model, like
$SO(10)$ GUT's or string models, where such Standard Model singlets as
$N_R$ with masses in the phenomenologically interesting range
(typically, $M_R \sim 10^{12}-10^{16} \: GeV$, which corresponds to a
light neutrino mass $m_\nu \sim 10^{-3}-10 \: eV$) can be present.
Assuming one $RH$ neutrino per family, the Dirac (Majorana) mass in
(\ref{eq:seesaw}) is replaced by a 3x3 matrix in family space ${\cal
M}_D$ (${\cal M}_M$), and the light neutrino spectrum is obtained by
diagonalizing the symmetric matrix:
\begin{eqnarray}
 {\cal M}_{\nu} & = &
	- {\cal M}_D \; {\cal M}^{-1}_M \; {\cal M}^T_D  \nonumber \\
 & = & R_{\nu} \; Diag (m_{\nu_1}, m_{\nu_2}, m_{\nu_3}) \; R^T_{\nu}
\end{eqnarray}
Note that the mixing angles relevant for neutrino oscillations are given
by the analog of the CKM matrix, which also involves the charged lepton
sector\footnote{The charged lepton mass matrix is in general not
hermitian, so it is diagonalized by two unitary matrices: $Diag(m_1,
m_2, m_3) = R^L_e {\cal M}_e R^{R \: \dagger}_e$.}:
\begin{equation}
  V_L = R^L_e R_{\nu}
\end{equation}

In general, the entries of both the Dirac and the Majorana matrices are
free parameters, and one has to choose a specific ansatz in order to
make any definite prediction in the neutrino sector. It is often assumed
that the Dirac mass matrix has the same structure than the up quark mass
matrix\footnote{This arises naturally in Standard  Model extensions with
a quark/lepton symmetry, like $SO(10)$.}: ${\cal M}_D \sim {\cal M}_U$.
For the Majorana matrix, however, no such simplifying assumption can be
done, and it is necessary to assume a specific form. It follows that the
neutrino spectrum of a given model depends on the ansatz that has been
chosen\footnote{For example, if there is no significant hierarchy
between the heavy Majorana masses, the light neutrino masses are
expected to scale as the up quark squared masses: $m_{\nu_i} \sim
m^2_{u_i}/M_R$.}, which is not very satisfactory.

Alternatively, one can try to derive the structures of the Dirac and the
Majorana matrices from a symmetry. This symmetry has to act in a
different way on the three neutrino families, otherwise the matrices
would be unconstrained. Such a symmetry is called a {\it family
symmetry}. This approach has proved to be successful in the quark
sector, where, following the original idea by Froggatt and Nielsen
\cite{FN}, several groups \cite{LNS,IR,BR,JS,DPS,BLR} have shown that an
abelian family symmetry can reproduce the observed mass and mixing
hierarchies.


\section{Neutrino mass models with a $U(1)$ family symmetry}
\indent

The class of models we consider are extensions of the Minimal
Supersymmetric Standard Model (MSSM) with: (i) a gauge group $SU(3)_C
\times SU(2)_L \times U(1)_Y \times U(1)_X$, where $U(1)_X$ is an
abelian family symmetry; (ii) a SM singlet field $\theta$ with
$X$-charge $X_\theta = -1$, which is used to break $U(1)_X$ and to
generate fermion masses; and (iii) three $RH$ neutrinos $\bar N_i$ ($i$
is a family index), in addition to the MSSM spectrum, which are needed
to generate neutrino masses by the seesaw mechanism. We require that the
family symmetry reproduce the experimental data on quarks and charged
leptons, which forces it to be anomalous \cite{BR,BLR}, and that its
anomalies be compensated for by an appropriate mechanism (the
Green-Schwarz mechanism\footnote{Remarkably enough, the observed fermion
mass hierarchy, through this mechanism, fixes the Weinberg angle
\cite{Ibanez} to its standard value at the unification scale, $\sin^2
\theta_W = 3/8$. This, together with the anomalous character of
$U(1)_X$, suggests a superstring origin to the model.}).

In the following, we concentrate on the neutrino sector. We denote the
lepton doublets by $L_i$ (with their $I_W = + 1/2$ components $\nu_i$),
and the right-handed neutrinos by $\bar N_i$, and their charges under
$U(1)_X$ respectively by $l_i$ and $n_i$. We note $h_1$ and $h_2$ the
$X$-charges of the two MSSM Higgs doublets $H_1$ and $H_2$.

\subsection{Dirac and Majorana matrices}
\indent

Let us show how the Dirac (${\cal M}_D$) and Majorana (${\cal M}_M$)
matrices are constrained by the family symmetry. Each Dirac mass term
$L_i \bar N_j H_2$ carries an $X$-charge $p_{ij} = l_i + n_j + h_2$. If
$p_{ij} \neq 0$, the coupling is forbidden by $U(1)_X$, and the
corresponding entry of ${\cal M}_D$ is zero. However, if the excess
charge $p_{ij}$ is positive, one can write non-renormalisable
interactions involving the chiral singlet $\theta$:
\begin{equation}
  L_i \bar N_j H_2 \left( \frac{\theta}{M} \right) ^{p_{ij}}
\end{equation}
where $M$ is a large scale characteristic of the underlying theory
(typically $M \sim M_{Planck}$ or $M_{GUT}$). When $\theta$ acquires a
vev, $U(1)_X$ is spontaneously broken and effective Dirac masses are
generated:
\begin{equation}
  ({\cal M}_D)_{ij} \sim v_2 \left( \frac{<\theta>}{M}
\right)^{\:p_{ij}} 
\end{equation} 
where $v_2 = <H_2>$. Since $U(1)_X$ is broken below the scale $M$,
$\epsilon \equiv <\theta>/M$ is a small parameter. Thus the Dirac matrix
obtained has a hierarchical structure\footnote{Remember that the first
motivation for introducing a family symmetry was to understand the quark
and charged lepton mass hierarchies.}, with the order of magnitude of
its entries fixed by their excess charges under $U(1)_X$.

The entries of the Majorana matrix ${\cal M}_M$ are generated in the
same way, with non-renormalizable interactions of the form:
\begin{eqnarray}
  M_R \: \bar N_i \bar N_j \left( \frac{\theta}{M} \right) ^{q_{ij}}
\end{eqnarray}
giving rise to effective Majorana masses
\begin{equation}
  ({\cal M}_M)_{ij} \sim M_R \left( \frac{<\theta>}{M}
\right)^{\:q_{ij}} 
\end{equation}
provided that $q_{ij} = n_i + n_j$ is a positive integer (otherwise
$({\cal M}_M)_{ij} = 0$). Thus, the light neutrino mass matrix ${\cal
M}_{\nu}  = - {\cal M}_D {\cal M}^{-1}_M {\cal M}^T_D$, and consequently
the neutrino masses and mixing angles, is determined by the charges of
the leptons under $U(1)_X$. {\em No particular ansatz for ${\cal M}_D$
nor ${\cal M}_M$ is required.} Note, however, that each of the entries
is determined only up to an arbitrary factor of order one by the family
symmetry.

Of course, there is a large variety of models, depending on the charges
one assigns to the lepton fields. Contrary to the quark charges, whose
possible values are strongly restricted by the experimental data on
quark masses and CKM angles, the lepton charges are poorly constrained.
In the following, we present  two classes of models. The first one leads
to a hierarchical mass spectrum \cite{DLLRS,GN,Montreal,BLR}, the second
one has two quasi-degenerate neutrinos with a large mixing angle
\cite{BLPR}.

\subsection{Model 1: hierarchical mass spectrum}
\indent

By analogy with the quark and charged lepton mass matrices, we start
from a matrix with only one coupling allowed by the family symmetry:
\begin{eqnarray}
  {\cal M}_\nu &  =  &  m_{\nu_3} \left( \begin{array}{ccc}
	0  &  0  &  0 \\
	0  &  0  &  0 \\
	0  &  0  &  1   \end{array}  \right)
  \label{eq:hierarchy}
\end{eqnarray}
The breaking of $U(1)_X$ fills in the zero entries with powers of the
small parameter $\epsilon$, leading to a hierarchical spectrum. Assuming
that (a) the $X$-charges of all mass terms are positive ($p_{ij} \geq
0$, $q_{ij} \geq 0$) and (b) the dominant entry of each mass matrix is
the (3,3) entry, one automatically obtains pattern (\ref{eq:hierarchy}).
The light neutrino masses are then:
\begin{equation}
m_{\nu_1} \sim {m_3^{\: 2} \over M_3} \;
	\epsilon^{\: 2 \, (l_1 - l_3)} \; \; \; \; \; \; \;
  m_{\nu_2} \sim \frac{m_3^{\: 2}}{M_3} \;
	\epsilon^{\: 2 \, (l_2 - l_3)} \; \; \; \; \; \; \;
  m_{\nu_3} \sim {m_3^{\: 2} \over M_3}
\label{eq:mass}
\end{equation}
The $\nu_\tau$ mass is given by the usual seesaw formula ($M_3$ is the
mass of the heaviest $RH$ neutrino, $m_3$ the largest Dirac mass),
whereas the other neutrino masses are suppressed relative to
$m_{\nu_{\tau}}$ by powers of the small breaking parameter $\epsilon$.
Note that the hierarchy depends only on the $X$-charges of the lepton
doublets $L_i$. The lepton mixing matrix is:
\begin{equation} 
V_L \sim  \left( \begin{array}{ccc}
1 & \epsilon^{\: |l_1 - l_2|} & 
  \epsilon^{\: |l_1 - l_3|}  \\
\epsilon^{\: |l_1 - l_2|} & 1 & 
  \epsilon^{\: |l_2 - l_3|}  \\
\epsilon^{\: |l_1 - l_3|} & 
  \epsilon^{\: |l_2 - l_3|} & 1  \end{array} \right)
\label{eq:mixing}
\end{equation}

This model has several remarkable features. First, the neutrino mass and
mixing hierarchies do {\it not} depend on the particular form of the
Majorana matrix. This is a great difference with most seesaw models. The
reason for this is that the dependences of ${\cal M}_D$ and ${\cal M}_M$
on the heavy neutrino charges compensate for each other in ${\cal
M}_{\nu}$. Secondly, the mass spectrum obtained is naturally
hierarchical\footnote{The possibility of mass degeneracies will be
discussed in the next section.}, without hierarchy inversion: $m_{\nu_e}
\ll m_{\nu_\mu} \ll m_{\nu_\tau}$. Finally, the mixing angles and the
mass ratios are related by:
\begin{equation}
  \sin^2 \theta_{ij} \sim {m_{\nu_i} \over m_{\nu_j}}
\label{eq:angle}
\end{equation}
These relations, which are common to numerous seesaw models, show that
small mixing angles are associated with mass hierarchies. They also
imply $V_{e \nu_\mu} V_{\mu \nu_\tau} \sim V_{e \nu_\tau}$ in lepton
charged current, in analogy with $V_{us} V_{cb} \sim V_{ub}$ in quark
charged current.

The experimental data on solar neutrinos and atmospheric neutrinos put
constraints on the parameters of the model. For example, if one wants to
explain simultaneously the solar neutrino deficit by MSW $\nu_e
\rightarrow \nu_\mu$ transitions, and the atmospheric neutrino anomaly
by $\nu_\mu \rightleftharpoons \nu_\tau$ oscillations, one must choose:
\begin{eqnarray}
  l_1 - l_3 = 3  &  &  l_2 - l_3 = 1
\end{eqnarray}
which leads to the following spectrum:
\begin{eqnarray}
  \left\{ \begin{array}{ll}
	m_{\nu_e} \sim 10^{-5} & eV  \\
	m_{\nu_\mu} \sim 5.10^{-3} & eV  \\  
	m_{\nu_\tau} \sim 0.1 & eV
  \end{array} \right.  & &  \left\{ \begin{array}{l}
	sin^2 \: 2 \theta_{e \mu} \simeq 2.10^{-3} - 4.10^{-2}  \\
	sin^2 \: 2 \theta_{\mu \tau} \simeq 5.10^{-2} - 0.8
  \end{array} \right.
\end{eqnarray}
The uncertainties in the mixing angles are due to the fact that the mass
matrix entries are determined only up to a factor of order one by the
family symmetry. It is quite difficult to obtain a large mixing angle
with a hierarchical spectrum [see (\ref{eq:angle})], as required by the
atmospheric neutrino data. Furthermore, the tau neutrino is too light to
be a good candidate for hot dark matter. However, if one ignores the
atmospheric neutrino problem, it is possible to obtain a cosmologically
relevant tau neutrino and to account for the solar neutrino deficit at
once.

\subsection{Model 2: quasi-degenerate neutrinos}
\indent

As suggested above, in the context of abelian family symmetries, large
mixing angles are naturally related to mass degeneracies. It thus seems
rather difficult for the previous model to account for the atmospheric
neutrino anomaly, or to accommodate the large angle branch of the MSW
effect. Yet mass degeneracies and large mixing angles are not excluded:
according to (\ref{eq:mass}) and (\ref{eq:mixing}), $l_2 = l_3$ yields
$m_{\nu_\mu} \sim m_{\nu_\tau}$ and $V_{\mu \nu_\tau} \sim V_{\tau
\nu_\mu} \sim 1$. But the presence of unconstrained factors of order one
in each entry of ${\cal M}_\nu$ can upset these formulae. Also, an
accurate mass degeneracy requires fine-tuning of these factors. This
leads us to consider another class of family symmetries, allowing for
two equal\footnote{The equality of the couplings in
(\ref{eq:degeneracy}) follows from the symmetry of Majorana mass terms.}
couplings, e.g.:
\begin{eqnarray}
  {\cal M}_\nu &  =  &  m_{\nu_3} \left( \begin{array}{ccc}
	0  &  0  &  0 \\
	0  &  0  &  1 \\
	0  &  1  &  0   \end{array}  \right)
  \label{eq:degeneracy}
\end{eqnarray}
Such a matrix has two degenerate eigenvalues\footnote{The relative sign
simply means that the mass eigenstates have opposite $CP$ parities.},
$m_{\nu_2} = - \: m_{\nu_3}$, with a maximal mixing angle, $\sin^2 2
\theta = 1$. This degeneracy is slightly lifted by the breaking of
$U(1)_X$. In order to reproduce pattern (\ref{eq:degeneracy}), we choose
the following assignment of lepton charges:
\begin{eqnarray}
  \left\{ \begin{array}{l}
	l_1 = l' > l \\ l_2 = - l_3 = l  \end{array}  \right.
  &  \mbox{and}  &  0 \leq n_1 < (-n_3) \leq l \leq n_2
\label{eq:assignment}
\end{eqnarray}
and, for simplicity, we assume $h_1=h_2=0$. The light neutrino mass
matrix is then:
\begin{equation}
  {\cal M}_\nu \sim m_{\nu_3} \left( \begin{array}{ccc}
    \epsilon^{\: 2l'} & \epsilon^{\: l'+l} & \epsilon^{\: l'-l} \\
    \epsilon^{\: l'+l} & \epsilon^{\: 2l} & 1 \\
    \epsilon^{\: l'-l} & 1 & \epsilon^{\: 2l}   \end{array}  \right)
 \label{eq:Mnu}
\end{equation}
As expected, the spectrum contains two heavy, strongly degenerate
neutrinos
\begin{eqnarray}
	m_{\nu_1} \sim \epsilon^{\: 2l'} & \ll &
	m_{\nu_2} \simeq - \: m_{\nu_3}
\end{eqnarray}
with a squared mass difference determined by the family symmetry:
\begin{equation}
  \Delta m^2_{23} \; \sim \; m^2_{\nu_3} \: \epsilon^{\: 2l}
\end{equation}
Taking into account the charged lepton sector, one obtains the lepton
mixing matrix:
\begin{equation}
  V_L \; \sim \; \left( \begin{array}{ccc}
	1 & \epsilon^{\: l'-l} & \epsilon^{\: l'+l}  \\
	\epsilon^{\: l'-l} & \; \; \: \frac{1}{\sqrt{2}} &
		\frac{1}{\sqrt{2}}  \\
	\epsilon^{\: l'-l} & - \frac{1}{\sqrt{2}} & \frac{1}{\sqrt{2}}
  \end{array}  \right)
\end{equation}
The quasi-degenerate neutrinos are almost maximally mixed, while the
third neutrino, which is lighter, has small mixings with the other ones.

Such a mass and mixing pattern can account for the hot dark matter of
the Universe, and simultaneously explain the atmospheric neutrino
deficit in terms of $\nu_\mu \rightleftharpoons \nu_\tau$ oscillations.
This requirement, together with the experimental limits on $\bar \nu_\mu
\rightleftharpoons \bar \nu_e$ oscillations and the charged lepton
masses, fixes the parameters of the model to be:
\begin{equation}
  l = 2 \; \; \; \; \; \; \;
  l' = 5 \; \; \; \; \; \; \;
  m_{\nu_3} = 2-3 \: eV
\end{equation}
Thus, the mu and tau neutrinos are in the relevant mass range for hot
dark matter:
\begin{eqnarray}
  m_{\nu_e} \sim (5-8).10^{-7} \: eV  &  &
	m_{\nu_\mu} \simeq m_{\nu_\tau} = 2-3 \: eV
\end{eqnarray}
and, being strongly degenerate, they can oscillate with the parameters
needed in order to solve the atmospheric neutrino problem:
\begin{eqnarray}
  \Delta m^2_{\mu \tau} \sim (0.9-2).10^{-2} \: eV^2  &  \mbox{and}  &
	\sin^2 2 \theta_{\mu \tau} \simeq 1
\end{eqnarray}
Furthermore, $\bar \nu_\mu \rightleftharpoons \bar \nu_e$ oscillations
are found to be in the domain of sensitivity of the LSND and KARMEN
experiments:
\begin{eqnarray}
  \Delta m^2_{e \mu} = 4-9 \: eV^2  &  \mbox{and}  &
	\sin^2 2 \theta_{e \mu} \sim 10^{-3}
\end{eqnarray}
Finally, let us note that the solar neutrino problem cannot be solved
by this model, unless a sterile neutrino $\nu_s$ is added.

\section{Conclusion}
  \indent

We have presented two seesaw models based on an abelian family symmetry,
the one leading to a hierarchical mass spectrum, the other yielding an
accurate mass degeneracy and a large mixing angle between the two
heaviest neutrinos. In such models, the neutrino masses and mixing
angles are determined in terms of the lepton charges under the family
symmetry. Also, the squared mass difference between quasi-degenerate
neutrinos is predicted. No ansatz for the Dirac nor the Majorana matrix
is needed.  Furthermore, the fact that the same symmetry is able to
explain the observed fermion mass hierarchy and simultaneously
constrains the neutrino spectrum sets an interesting connection between
two fundamental problems in particle physics. Unfortunately, the lepton
charges, though constrained by experimental data, are not fully
determined by the model, which leads us to consider different classes of
models, corresponding to different mass patterns. However, some generic
properties of abelian family symmetries suggest that the model may
originate from a more fundamental theory, which would fix all of its
parameters.

\vskip 1cm
\noindent
{\bf Acknowledgements}
\vskip .5cm

This talk is based on work done in collaboration with P. Bin\'etruy and
P. Ramond, and with P. Bin\'etruy, S. Petcov and P. Ramond.


\end{document}